\documentclass[twocolumn,floatfix,pra,aps]{revtex4-1}
\usepackage{graphicx,amssymb}

\begin{document}

\newcommand{\Er}{Er$^{3+}$:Y$_{2}$SiO$_{5}$ }
\newcommand{\levela}{$^{4}$I$_{15/2}$ }
\newcommand{\levelb}{$^{4}$I$_{13/2}$ }
\newcommand{\system}{$\Lambda$-system }

\title{Approaches for a quantum memory at telecommunication wavelengths}

\author{Bj\"orn Lauritzen}
\email{bjorn.lauritzen@unige.ch}
\author{Ji\v{r}\'i Min\'{a}\v{r}}
\author{Hugues de Riedmatten}
\altaffiliation{Current addresses: 1) ICFO-Institute of Photonic Sciences, Mediterranean Technology Park, 08860 Castelldefels (Barcelona), Spain 2) ICREA-Institució Catalana de Recerca i Estudis Avançats, 08015 Barcelona, Spain}
\author{Mikael Afzelius}
\author{Nicolas Gisin}

\affiliation{Group of Applied Physics, University of Geneva, CH-1211 Geneva 4, Switzerland}

\begin{abstract}

We report experimental storage and retrieval of weak coherent
states of light at telecommunication wavelengths using erbium ions doped into a
solid. We use two photon echo based quantum storage protocols.  The first
one is based on controlled reversible inhomogeneous broadening
(CRIB). It allows the retrieval of the light on demand by
controlling the collective atomic coherence with an external
electric field, via the linear Stark effect. We study how atoms in the excited state affect the signal to noise ratio of the CRIB
memory. Additionally we show how CRIB can be
used to modify the temporal width of the retrieved light pulse.  The second protocol is based on atomic frequency combs
(AFC). Using this protocol we also verify that the reversible mapping is phase preserving
by performing an interference experiment with a local oscillator.
These measurements are enabling steps towards solid state quantum
memories at telecommunication wavelengths. We also give an outlook on possible improvements.

\end{abstract}

\date{\today}
\maketitle
\section{Introduction}
Light-matter interfaces in general, and quantum memories for single photons in particular, are of great interest for the
field of quantum information \cite{Hammerer2008}.
Quantum memories could be used to synchronize probabilistic quantum processes. These are essential ingredients in, for instance, quantum repeater architectures which would allow us to extend fiber-based quantum communication
schemes beyond today's  distance limit set by attenuation in optical
fibers \cite{Briegel1998,Duan2001,Sangouard2009a}.
Another potential application is to transform a probabilistic, but heralded, single photon source into a close to deterministic source of single photons, which is of interest for a variety of experiments and applications \cite{Lounis2000}.

A photonic quantum memory requires reversible
mapping of light, at the single photon level, onto long-lived coherent atomic
excitations.
Important progress has been made during the last years towards
the realization of an efficient and coherent quantum memory
\cite{Chaneliere2005,Eisaman2005,Choi2008,Hetet2008,Riedmatten2008,Hedges2010,Reim2010}.
Almost all of these experiments have been performed with atomic
transitions in the visible range. However, in the context of
quantum communication, it would be of great interest to have the ability
to store and retrieve photons at telecommunication wavelengths
(around 1550 nanometers), which can be transmitted with low loss in
optical fibers. Such a quantum memory could be easily integrated
into fiber optical networks. It would also be useful to realize a
narrow band deterministic source of single photons at
telecommunication wavelengths for quantum
communication. Finally, a quantum memory capable of storing
photons in this range is also required for some
efficient quantum repeater architectures. \cite{Sangouard2009a, Sangouard2007a, Sangouard2008}

One possibility to interface photons at telecommunication wavelengths
and quantum memories is to use frequency conversion techniques, i.e. to
convert the wavelength of the telecom photon to the resonance
frequency of existing solid state or atomic gases quantum memories
in the visible range or vice versa \cite{Tanzilli2005,Takesue2010,Curtz2010}. But such a conversion will induce
additional loss and increase the complexity of the experiment.
Moreover, the application of a strong pump laser applied to a
optically nonlinear medium, a commonly used method for such a conversion, can lead to noise
which may blur the single photon signal \cite{Thew2006}.
Another possibility is to directly use atomic systems with an
optical transition in the telecommunication range. A promising candidate in this context is erbium doped into crystals, since erbium has a well-known transition around 1530 nm. In particular Er$^{3+}$ ions doped into Y$_2$SiO$_5$ crystals
\cite{Bottger2006} are highly interesting since an extremely long optical coherence time of around 6 milliseconds has been obtained is this solid-state system \cite{Sun2002,Bottger2006a}.

Quantum storage in erbium doped solids is challenging because of limitations in
the required memory preparation as described in \cite{Lauritzen2008} for \Er.
Here we will report on another difficulty when implementing  a quantum memory with this material.
Residual population in the excited state due to imperfect memory preparation together with a low memory efficiency causes a low signal-to-noise ratio when performing storage experiments at the single photon level. We will discuss this issue in more detail and present solutions to this problem.

Despite these challenges, we report proof of principle demonstrations of
photon echo based storage at the single photon level in an erbium
doped crystal. The first one is based on controlled reversible inhomogeneous broadening (CRIB)
\cite{Moiseev2001,Nilsson2005,Kraus2006,Alexander2006}. The main results of this work
have already been published elsewhere \cite{Lauritzen2010}.
Here we give a more detailed study on how the weak CRIB signal
is affected by the fluorescence noise
from excited atoms. We also show how this technique can be used to
control the temporal profile of the retrieved pulses \cite{Moiseev2010}.
The second method is based on an atomic frequency comb (AFC) \cite{Afzelius2009a}
written into the inhomogeneously broadened absorption
profile of the rare earth ion doped solid. It has already been
implemented in a range of different materials at different wavelengths
\cite{Riedmatten2008,chaneliere-2009,Afzelius2010,Sabooni2010,Bonarota2010,Amari2010,Bonarota2010a,Clausen2010,Saglamyurek2010}.
Here we present the first results at telecommunication wavelengths.
Besides the novelty of the material and the wavelength, this
experiment also includes previously unexplored aspects of the AFC
protocol, such as an alternative way of creating the atomic
frequency comb using standard hole burning techniques. We also
study the coherence of the storage process by performing an
interference experiment with a local oscillator.
Even if the efficiencies obtained for both protocols  are still low,
it shows the feasibility of a memory at telecommunication wavelengths
and it hopefully will stimulate further research on the spectral
properties of erbium doped solids.

The article is structured as follows. In section
\ref{photonecho}, we briefly review the theoretical basis and the
experimental state of the art for the two quantum storage
protocols used in this paper.
Section \ref{experimental} describes the experimental procedure
used. In section \ref{CRIB} and \ref{AFC}, we present the results
obtained for the AFC and CRIB protocols, respectively. Finally, in section
\ref{combination} we propose to combine the two methods and show
first experimental results on that.

\section{Photon echo based quantum storage}
\label{photonecho}
In this section we discuss the underlying process
of both photon echo based protocols we explored in this work.
For this let us consider  an ensemble of atoms forming
an inhomogeneously broadened atomic absorption. If
a single photon spectrally matching this broadened line gets
absorbed, the atoms will be in a superposition of states
(collective excitation):
\begin{equation}
|\psi\rangle=\frac{1}{\sqrt{N}}\sum_{j=1}^{N}e^{i2\pi\delta_{j}t}e^{ikz_j}|g_{1}\ldots
e_{j}\ldots g_{N}\rangle. \label{collstate}
\end{equation}
$g_{j}$ denotes atom j in the ground, $e_{j}$ in the excited
state. A dephasing will take place since atoms absorbing at
different frequencies within this line will acquire different
phases $\delta_{j}t$, where $\delta_{j}$ denotes the detuning from the central absorption line. This dephasing
inhibits the collective emission of light by the ensemble.
However, if one finds a way to undo this dephasing, i.e. to obtain
an overall phase factor of one after a time $T$ for all atoms,
such an emission will take place. In the classical two and three
pulse photon echo, this is done by the application of bright
optical pulses on the optical transition. These methods are
interesting since they enable the storage and retrieval of large
trains of classical optical pulses \cite{Lin1995}. However, conventional
photon echo techniques have strong limitations regarding the
storage of single photons \cite{Ruggiero2009, Sangouard2010}.
The main reason is the application of the the strong pulses
that are necessary to induce the rephasing.
They will transfer population from the ground- to the excited state and
therefore cause an intrinsically low fidelity due to noise
from incoherent de-excitation (fluorescence).

In order to use photon echo techniques as quantum storage
protocols, it is thus necessary to devise methods where the
rephasing mechanism does not involve a population of
the excited state. In this article we present measurements on
photon echoes at the single photon level using two different
schemes fulfilling this requirement: controlled reversible inhomogeneous broadening (CRIB) and
atomic frequency combs (AFC), which we are both going to describe in
more detail now.

\subsection{Controlled Reversible Inhomogeneous Broadening}
The underlying rephasing mechanism of this scheme is based on a
reversal of the inhomogeneous broadening in a controlled way
(therefore it is known as \textit{controlled reversible
inhomogeneous broadening}, CRIB). This protocol was first proposed
for hot atomic gases, where the Doppler shift of atoms depends on
the direction of laser beams \cite{Moiseev2001}. CRIB was later
extended to rare-earth doped solids by three different groups
\cite{Nilsson2005,Kraus2006,Alexander2006}. The idea is the
following \cite{Tittel2009a,Nilsson2005}: First, a narrow absorption line is created within a
large transparency window. This can be done using optical
pumping techniques as will be described in section \ref{spectral}. This initially narrow
absorption line is then artificially broadened. In our case this is
done by the application of an electric field gradient in the
direction of light propagation ($z$). Since the ions possess a
permanent electric dipole moment, their optical resonance
frequency will shift due to the linear Stark effect by an amount $\delta_j$
that depends on the position $z_{j}$ of ion $j$.
After the absorption of a photon the atoms will be in a state as given by equation \ref{collstate} and the atomic dipoles will dephase accordingly.
This dephasing can be reversed by changing the sign of the detuning
$\delta_{j} \rightarrow -\delta_{j}$ at a time $\tau$, which can be
done by simply flipping the polarity of the electric field. Note that $\tau$ can be chosen after the absorption
(\textquotedblleft on demand\textquotedblright readout of the memory).
The state of the system at a time t after the flip becomes
\begin{equation}
|\psi>=\frac{1}{\sqrt{N}}\sum_{j=1}^{N}e^{i2\pi\delta_{j}\tau}e^{-i2\pi\delta_{j}t}e^{ikz}|g_{1}\ldots
e_{j}\ldots g_{N}>. \label{rephasing_CRIB}
\end{equation}
It is obvious that at the time $t=\tau$, i.e. a total time
$2\tau$, the spectral phases will have canceled. The
atomic dipoles are back in phase and the ensemble will
collectively emit an echo of the incident light in the forward direction.

The forward retrieval efficiency is given by \cite{Sangouard2007}
\begin{equation}
 \eta_{CRIB}=d_{br}^{2}e^{-d_{br}}e^{-d_{0}}e^{-t^2\tilde{\gamma}^2},
\label{CRIB_efficiency}
\end{equation}
where $d_{br}$ is the optical depth of the broadened peak, and $\tilde{\gamma}=2\pi\gamma$ the spectral width of the initial peak (standard deviation). The first term gives the raw efficiency for the photon to be absorbed and re-emitted. The second term accounts for the possibility for the photon to be re-absorbed by the medium. In case of inefficient optical pumping, (re-)absorption of the light by an absorbing background with optical depth $d_{0}$ occurs, which is taken into account here as well. The last term describes the decoherence due to the finite width of the initial peak. Note that a unit efficiency memory can be achieved using a backward readout \cite{Sangouard2007}, provided that $d_{0}=0$.

CRIB was first realized in europium doped
Y$_{2}$SiO$_{5}$ crystals \cite{Alexander2006}.
Since the initial demonstration, a lot of progress has been made
concerning its efficiency using praseodymium as dopant, approaching $70\%$ in the quantum regime \cite{Hedges2010}. We have recently demonstrated the
feasibility of the protocol at the single photon level in the telecommunication band using an erbium doped sample \cite{Lauritzen2010}. In this article we present
an analysis of the noise occurring in this experiment and
the methods we applied to reduce it which allowed us to work in this regime.
Furthermore we present measurements showing pulse-compression
and -stretching at the few photon level.
Note that CRIB has also been realized with controlled broadening of spin
transitions in a rubidium vapor \cite{Hetet2008a}. Its capability
to serve as a pulse sequencer and pulse shaper also was shown
experimentally \cite{Hosseini2009}.

\subsection{Atomic Frequency Combs}
The rephasing mechanism in this scheme is based on a periodicity
in the absorption profile \cite{Afzelius2009a}. We  start with a number of
equidistant absorption peaks with a separation of $\Delta$ (atomic
frequency comb, AFC).  Let us consider an incident photon with a
central frequency $\omega_{0}$ and a spectral width that is
smaller than the total width of the comb. The light will be
absorbed as a single excitation distributed over the atoms forming
the absorption peaks.

 We can write the state of the atoms in the same manner as in Eq.
\ref{collstate}. In the ideal case of very sharp comb peaks $\delta_{j}$ is given by
$\delta_{j}=n_{j}\Delta$, where $n_{j}\in{\bf Z}$ is the number of
the peak the ion belongs to.  It
is easy to see that, due to this periodicity, for times
\begin{equation}
T_{m}=\frac{m}{\Delta} \label{rephasingtime}
\end{equation}
($m \in{\bf N}$) the phase factors all are equal to one, i.e. the
atomic dipoles are back in phase and a collective emission will
take place (note that here the unit of  $\Delta$ is Hertz).
Unlike in the CRIB protocol the rephasing mechanism is
an intrinsic property of the spectral shape and no external
manipulation is required. However, for this reason the moment of
rephasing can not be changed after the absorption. An
additional coherent transfer of the excitation to a ground state
spin level and back is required \cite{Afzelius2009a} in order to
improve the scheme from a delay to an on demand memory.

The efficiency of the AFC scheme in the forward direction for the first echo is given
by \cite{Afzelius2009a, Riedmatten2008}:
\begin{equation}
\eta_{AFC}=\frac{d^2}{F^2}e^{-\frac{d}{F}}e^{-d_{0}}e^{-\frac{1}{F^2}\frac{\pi^{2}}{4\ln2}}.
\label{AFCeq}
\end{equation}

The finesse $F$ is the ratio between the width of the peaks $\gamma$ and their separation $\Delta$.
$d$ is the optical depth of the peaks. Note that this formula is similar to the one given for the CRIB scheme (eq. \ref{CRIB_efficiency}). The optical depth of the broadened peak is simply replaced by the effective optical depth  of the AFC, $d/F$. If one writes the finesse $F$ in terms of the storage time $T_{1}=1/\Delta$ and linewidth $\tilde{\gamma}$, one will find that the last term corresponds to the decoherence term in equation \ref{CRIB_efficiency}. Also here an absorbing background is taken into account. Note that, as for CRIB, unit efficiency is possible in backward configuration \cite{Afzelius2009a}.

The AFC protocol has been used to demonstrate the first solid
state light matter interface at the single photon level
\cite{Riedmatten2008} and more recently to demonstrate entanglement between a photonic qubit and an atomic excitation in a solid \cite{Saglamyurek2010,Clausen2010}.  The extension of the scheme by spin wave
storage was recently reported \cite{Afzelius2010}. Further work has shown improved storage efficiencies \cite{Bonarota2010,Sabooni2010,Amari2010}. The
most important feature of this protocol, that one can efficiently store  and
retrieve multiple temporal modes (see sec. \ref{comparison_mult}), was also shown
\cite{Riedmatten2008,Usmani2010,Bonarota2010}.
In this work we demonstrate AFC experiments at telecommunication wavelengths in an erbium doped crystal.
Furthermore we show how the phase of the AFC echo can be shifted by changing
the central frequency of the AFC. We obtain interference fringes
with a visibility of 90\%. Additionally, we carried out the same
experiment with the second AFC echo ($m=2$). Also here we observe interference fringes.
We find that the phase of these fringes is twice as large as for the fringes of the first echo.

\subsection{Comparison of the two protocols}

At this point we would like to give a brief comparison of the two protocols.

\subsubsection{Multimode Capacity}
\label{comparison_mult}
The possibility to store a train of incident photons (multiple temporal modes) in a quantum memory is of great interest in the context of quantum repeater architectures \cite{Simon2007}. It has been shown that the number of temporal modes $N$ that can be efficiently stored in a CRIB memory grows linearly with the optical depth $d$ of the initial unbroadened absorption line  \cite{Simon2007,Nunn2008}. This can be understood by the following arguments:

The number of temporal modes that can be stored in the memory is roughly given by the ratio of the time $\tau$ (from the first input pulse to the moment of switching) to the duration of the each input mode $T_{CRIB}$: $N\propto\frac{\tau}{T_{CRIB}}$. The storage time of the CRIB protocol is of the order of the inverse of the width of the initial peak $\tau\propto\frac{1}{\gamma_{CRIB}}$. On the other hand, the broadened absorption profile should be at least as large as the linewidth of the incident light (see section \ref{CRIB}).  The number of temporal modes that can be stored is thus obviously given by the factor with which the initial line was broadened. The optical depth $d_{br}$ of the broadened line, however, decreases linearly with this factor and this diminishes the storage efficiency (see eq. \ref{CRIB_efficiency}). In short, to double the number of temporal modes that can be stored while keeping the storage efficiency constant, one has to double the initial optical depth $d$.

The situation is different for the AFC protocol. In order to increase the spectral width of the comb one can add peaks to it.  In fact, the number of modes is limited by the number of peaks \cite{Afzelius2009a}. The optical depth and the finesse are not affected and the efficiency does not change (eq. \ref{AFCeq}). The limitations for the spectral width (and thus the number of peaks) of a comb, are material properties. These are for example the inhomogeneous broadening as well as hyperfine level spacings.
The number of temporal modes that can be stored does not depend on the available optical depth, contrary to all other QM proposals.
Furthermore is the efficiency for all modes the same. This is in contrast to the CRIB protocol where the efficiency decays exponentially.

An interesting aspect in the context of the multimode capacity of the two protocols is that AFC is a First In - First Out (FIFO) memory \cite{Usmani2010, Riedmatten2008}, while CRIB is a First In - Last Out (FILO) memory \cite{Hosseini2009}. This may be of interest for future applications of quantum memories.

\subsubsection{Spin Wave Storage}
\label{Spin Wave Storage}
The transfer of the excitation to a long-lived ground state level using control pulses ($\pi$-pulses) is a required step for an AFC memory in order to achieve an on-demand readout. However, it can also be used in the CRIB scheme to increase the storage time \cite{Nilsson2005, Kraus2006}. In the case of a multimode memory discussed before, the bandwidths of the control pulses have to be sufficiently large in order to transfer all ions of the collective ensemble efficiently. In order to achieve this, one may use chirped pulses such as complex hyperbolic secant pulses \cite{Seze2003, Rippe2005, Rippe2008, Afzelius2010, Minar2010}. These require a sufficiently long time between input mode and echo - in order to be able to apply them over the entire bandwidth - as well as a high oscillator strength of the transition. To realize an efficient multimode AFC-memory with spin-wave storage can thus be difficult.
Since in CRIB the width of the spectrum can be controlled externally, one can narrow the required bandwidth to the width of the initial peak. Note that in this case  the electrical control field is off during the transfer.
The ability of being able to fully control the excitation
spectrum with the electric field is a great advantage here, since
it considerably relaxes the constraints on the control pulses to
obtain a efficient transfer. Thus an efficient transfer is more easily achievable for the CRIB scheme. Note that both protocols require a third ground state level for spin wave storage.

\subsubsection{Material Aspects}
The CRIB protocol requires the existence of a linear Stark or Zeeman effect in the medium which allows to induce and control the inhomogeneous broadening. In thulium doped YAG for example, the Zeeman effect is very weak since Tm$^{3+}$ is not a Kramers ion and the magnetic field only acts on the nuclear spin. A fast switching of high magnetic fields, as they would be required in this material, is not possible. Moreover, a linear Stark does not exist in Tm$^{3+}$:YAG due to the crystal symmetry. While it is thus impossible to use Tm$^{3+}$YAG in CRIB experiments, AFC experiments do not require either of the two and have already successfully been achieved in this material \cite{chaneliere-2009,Bonarota2010,Bonarota2010a}. AFC can thus be implemented in a larger range of materials.

To sum up: the main advantage of the AFC is the multimode capacity.  It can be performed in a higher variety of materials since it does not depend on the existence and the strength of the interaction with an external electric or magnetic control field. However, the transfer of excitations to a long-lived ground state level  required for an on-demand readout in a AFC memory is more challenging. Both protocols have their advantages and are worth being investigated.



\section{Experimental Setup and Noise Reduction}
\label{experimental} In this section we describe the experimental
setup as well as the optical pumping techniques and the
experimental cycle common to both experiments presented in this
work. We also study the main source of noise present in the
system, which could be identified as fluorescence due to imperfect memory preparation.
We show how this noise can be reduced.
\subsection{Setup}
\begin{figure}[htb]
\includegraphics[width=\columnwidth]{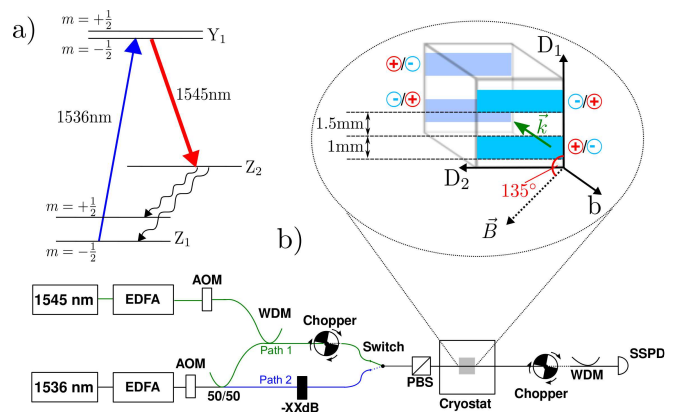}
\caption{(COLOR ONLINE) a) Level scheme of the material.
Z$_{1}$ and Z$_{2}$ denote the first and second crystal field level of the ground state, Y$_{1}$ the first crystal field level of the excited state.
b) Experimental setup: Both lasers are amplified using erbium doped fiber amplifiers (EDFAs).
The pump laser (1536 nanometers) is split into two paths.
For state preparation (path 1) it is combined
with the stimulation laser (1545 nanometers) using a wave division
multiplexer (WDM). In path 2 it is attenuated to produce
the weak optical pulses to store in the medium.  An
optical switch allows to send either of them into the sample.  The
strong path is blocked before the cryostat during the measurement
sequence and after the cryostat during state preparation,
respectively, in order to protect the superconducting single
photon detector (SSPD) and to avoid noise from a leakage in the
optical switch. Pulses are created with acousto-optical modulators
(AOMs). Two AOMs were necessary for the laser at 1536nm in
order to obtain a sufficient extinction
ratio. The polarizing beam splitter (PBS) is set parallel to the
D$_{2}$-axis to achieve maximal optical depth  and polarization is
adjusted before to maximize the power arriving at the sample. Inset:
Illustration of the crystal with electrodes, magnetic field and
direction of light propagation indicated.}\label{setup}
\end{figure}

The sample used is an Er$^{3+}$:Y$_{2}$SiO$_{5}$ crystal (10ppm)
placed on the cold finger of a pulse tube cooler (Oxford
Instruments), cooled down to 2.6K. The Y$_{2}$SiO$_{5}$
crystal has three mutually perpendicular optical-extinction axes
labeled D1, D2, and b. Its dimensions are $ 3.5 \times 4 \times
6$mm along these axes. The Er$^{3+}$ can be found at two crystallographic inequivalent sites (sites 1 and 2).
All measurements in this work were carried out on ions at site 1.
A magnetic field was used
to induce a Zeeman splitting necessary for state preparation.
It was provided by a permanent magnet outside the cryostat and was applied in the
$D_{1}-D_{2}$ plane at an angle of $\theta=135^{\circ}$ with
respect to the $D_{1}$-axis \cite{Hastings-Simon2008} (Figure
\ref{setup}, inset). For all measurements it was set to 1.5 mT in
order to allow for an optimal optical pumping efficiency and not
to decrease the optical depth by separating the inhomogeneously
broadened Zeeman transitions completely. The light
was traveling in a direction parallel to b and its polarization
was aligned to maximize the absorption \cite{Bottger2006}. The
optical depth of the sample was 2. The electric quadrupole field
for the CRIB experiment was applied with the use of four
electrodes placed on the crystal as shown in the inset of figure
\ref{setup} and described in \cite{Minar2009a}.

\begin{figure}[htb]
 \includegraphics[width=\columnwidth]{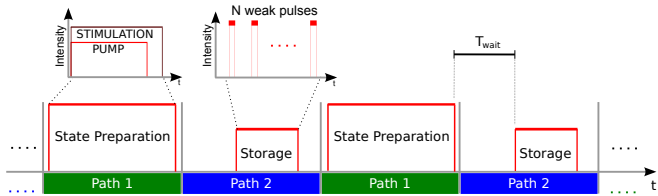}
\caption{(COLOR ONLINE) Pulse sequences used in the experiments. During state
preparation ions are pumped from one to the other ground state
Zeeman level via the excited state as described in the text and
illustrated in \ref{setup}a. The stimulation laser is still left
on for 23.5 milliseconds after the optical pumping in order to deplete the excited state.
After state preparation the switch is set to the attenuated path 2 and a series of
measurements is carried out. A number of $N$ (typically $N=8000$)
consecutive weak optical pulses is sent into the sample. For each
of these trials the time-to-analog-converter is triggered and in
case of a detection a count is added to the measured time bin in a
histogram.  The sequence is repeated at a rate of 3Hz.}
\label{pulse}
\end{figure}

The experimental setup is shown in figure \ref{setup}b. The
experimental sequence was divided into two parts: the preparation
of the memory and the storage and retrieval of the weak pulses.
For state preparation an external cavity diode laser at 1536
nanometers (Toptica) was used. It also served as light source for the weak input pulses.
The timing of the experimental sequence is shown in Fig.
\ref{pulse}. The preparation of the memory consisted in preparing
the specific absorption structure required for each protocol (a
single absorption line for CRIB and a comb for the AFC). The
preparation was done with optical pumping techniques (see section \ref{spectral})
and lasted 120 milliseconds. After the preparation, we waited a time
$T_{wait}$ to allow the atoms in the excited state to decay to
the ground state before the actual storage experiment.  For measurements
at the single photon level, two optical paths were necessary: one for the intense pulses used in the
optical pumping  (strong path) and one for the weak pulses at the single photon
level to be stored and retrieved in the crystal (weak path). In the weak
path, the light was attenuated to the single photon level
with a variable fiber attenuator. The two paths were recombined with
a fiber coupled optical switch. The weak output mode was detected
using a superconducting single photon detector (SSPD)
\cite{Goltsman2001} run in constant voltage mode with an
efficiency of $\sim7$\% and a dark count rate of $10\pm5$ Hz. During state preparation the SSPD
was blocked with a mechanical chopper in order to avoid it from being blinded by the relatively strong preparation pulses. During the
storage sequence, the leakage from the bright arm was further
blocked with another mechanical chopper in order to prevent noise.

\subsection{Spectral Tailoring}
\label{spectral}
 Both protocols require spectral tailoring of the
inhomogeneous absorption profile. This can be done by optically pumping ions from one
to another ground state level via the excited state. For this
reason the degeneracy of the ground state levels, as it is present
in \Er, has to be lifted by the application of an external
magnetic field making use of the Zeeman effect. 

The optical pumping efficiency depends on the ratio between the
ground state relaxation time $T_Z$ and the excited state
relaxation time $T_1$, and on the branching ratio $\beta$ of the two
transitions connecting the ground state levels to the excited state. In
erbium doped Y$_{2}$SiO$_{5}$, $T_1$ is about 11 milliseconds \cite{Bottger2006} and $T_Z$ has been
measured to reach up to 130 milliseconds for similar experimental conditions
\cite{Hastings-Simon2008a}. $T_Z$ and $\beta$ are strongly dependent on the
direction of the magnetic field with respect to the crystal axes.
T$_{Z}$ also strongly depends on the temperature of the sample.
The branching ratio for the optimal $T_Z\approx$130 ms we found is of order of $\beta$=0.1. This means
$\sim90$\% of the population de-excites down to the initial state.
 Under these conditions, the number of optical pumping cycles that can be
performed is low, and it is very difficult to efficiently
transfer population from one ground state level to the other. The
optical pumping efficiency can be enhanced by artificially
decreasing $T_1$ using stimulated emission to a short lived
auxiliary level that quickly decays to the ground state via a
non-radiative decay \cite{Lauritzen2008}. Hence, two lasers are
used for  the optical pumping, a pumping laser at 1536 nanometers, and a
stimulation laser at 1545 nanometers.

\begin{figure}[htb]
\includegraphics[width=\columnwidth]{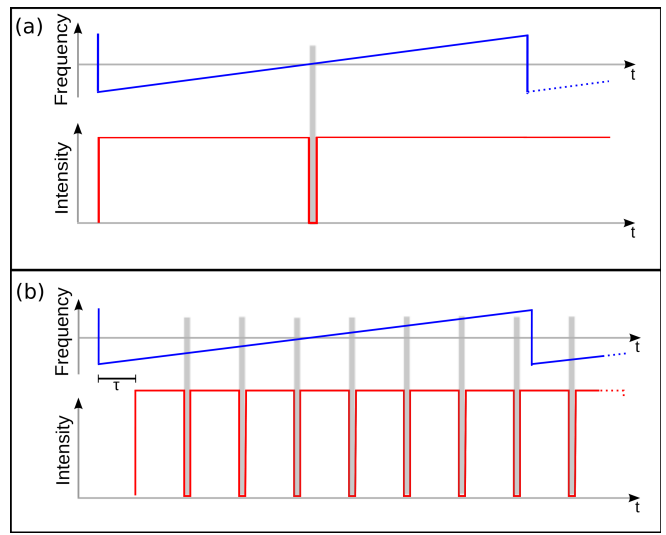}
\caption{(COLOR ONLINE) Pulse sequences used for the preparation. To create a spectral pit the pump laser is repeatedly swept.
To create the central peak required for the CRIB protocol it is turned off each time it passes the central frequency (a).
To create a comb structure it is turned off periodically. The trigger delay $\tau$ between sweep- and pulse-sequence allows to to create combs that are shifted in frequency and was used to perform the measurements that show the preservation of coherence during storage (see section \ref{AFC} and fig. \ref{interference}). } \label{pulse_sequence}
\end{figure}

The experimental cycle is illustrated in figure
\ref{pulse}. Each preparation sequence takes 120 milliseconds of
optical pumping during which both lasers are sent into the sample.
The frequency of the pump laser is repeatedly swept. In this way a
wide pit is burned into the inhomogeneously broadened absorption
line at 1536 nanometers. If the light is amplitude modulated using an
acousto-optical modulator (AOM) and turned off each time the laser passes a
frequency $\omega_{gate}$ a narrow absorption feature will be left
at this frequency (fig. \ref{pulse_sequence}a). For preparation of the frequency comb the laser
has to be periodically modulated as indicated in figure
\ref{pulse_sequence}b.

\subsection{Storage and retrieval experiments}
The time available to perform the actual storage experiments is limited by
the Zeeman lifetime of $T_{Z}\lesssim 130$ milliseconds \cite{Hastings-Simon2008a}
of the material. Population left in the excited state after preparation will lead to noise from fluorescence
during the storage experiments.
In order to deplete the excited state, the laser at 1545 nanometers is left on for an additional time
$T_{extra}$ after the AOM of the pump laser has been closed completely (figs. \ref{setup} and \ref{pulse}).
During the storage part the preparation path is closed and the chopper of the detection path is opened at the same
time. At a rate of 200 kHz  a sequence of $N$ strongly attenuated
pulses is sent into the sample beginning at a time $T_{wait}$
after the pump pulse. At the beginning of each of these trials a start signal
is sent  to a time to analog converter (TAC). A detection with
the SSPD provides the stop signal. The incident pulses are
weak coherent states of light $|\alpha\rangle_{L}$ with a mean
number of photons $\overline{n}=|\alpha|^{2}$. We determined
$\overline{n}$ by tuning the frequency of the laser out of the
spectral region of the inhomogeneously broadened absorption line
and measuring the number of photons arriving at the SSPD. By
measuring the losses between the input of the cryostat and the
detector (transmission$\approx15\%$) and the efficiency of the latter, we could trace back the average
number of photons per pulse arriving at the crystal. The results were checked
with a photo diode by a measuring the intensity of the same pulses
without attenuation. Measurements were performed at
$T_{wait}=86$ ms after state preparation in order to reduce noise
from from fluorescence as explained in section \ref{CRIB}.

\subsection{Noise reduction}
\label{noisereduction}
An experimental issue arising when input pulses are at the single
photon level is the fluorescence from atoms left in the excited
state due to an imperfect preparation of the memory. As described
above, in order to create the initial narrow absorption line, a
population transfer between two ground states ( in our case Zeeman
states) using optical pumping via the excited state is used. In
case of an incomplete transfer there will be population remaining
in the excited state after the preparation sequence. If the
depletion of this level is slow, which is the case for rare-earth
ions, this can lead to a high noise level from fluorescence that
will blur the weak echo pulse. This problem is especially important
for Erbium doped Y$_{2}$SiO$_{5}$, were the optical relaxation time is very
long ($T_1$ = 11 milliseconds \cite{Bottger2006}).
In order to investigate this issue, we performed an experimental
charaterization of the noise in our system.

\begin{figure}[htb]
\includegraphics[width=240pt]{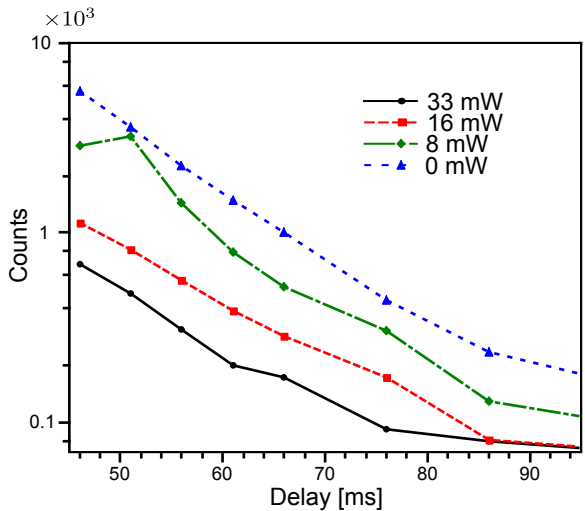}
\caption{(COLOR ONLINE) Noise measurement: Number of counts as
function of the delay after state preparation for different power
of the stimulation laser (logarithmic plot).
An exponetial fit to the data gave a decay time of 11 milliseconds which confirms that
the noise is due to ions left in the excited state.
Application of the stimulation laser yields a strong reduction of this noise. Integration time
was 100 seconds and $N=8000$ for each data point. $T_{extra}=10$ms for all measurements. The power of the stimulation laser was measured before the cryostat. }\label{noise}
\end{figure}

For this experiment we performed the state preparation as described above, leaving a single
absorption line at the center of the spectral pit. We then measured the number of counts
in the same temporal mode as we would do it for a
storage experiments, but with $\bar{n}=0$, i.e. no photons at the
input. We carried out experiments for different powers of the stimulation laser at 1545 nanometers.
For all cases the stimulation laser was left on for a time $T_{extra}$=10ms after the pump laser was switched off.
The result is presented in figure \ref{noise}.
One can see that the noise counts diminish
with a decay time of 11 milliseconds. This confirms that this noise is coming
from fluorescence due to ions left in the excited state after
memory preparation. Furthermore, the application of the stimulation laser helps
to reduce the noise significantly. However, with the limited
stimulation power available, we did not manage to quench it
completely. In order to perform experiments at the single photon
level it is thus advisable to go to high waiting times also with
stimulation laser applied. A drawback will be, that the memory efficiency goes down
since the created absorption structure decays with $T_{Z}$ which is only one order of magnitude longer
than the excited state lifetime $T_{1}$.
One thus has to find a compromise between storage efficiency and noise level. In section \ref{CRIB} we
show an analysis of the CRIB memory with respect to the efficiency
and the signal to noise ratio for different delays. Note that this
analysis also holds for the AFC measurements since they have been
performed under the same conditions.
\section{Controlled Reversible Inhomogeneous Broadening}\label{CRIB}
\begin{figure}[htb]
\includegraphics[width=214pt]{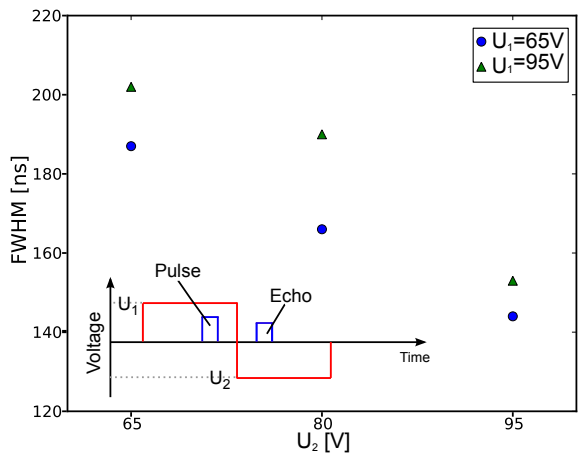}
\caption{(COLOR ONLINE) Full width at half maximum (FWHM) of
the CRIB echo for three different voltages $U_{2}$
applied to the sample after switching the polarity of the electrical field
as sketched in the inset.
Before switching a voltage of $U_{1}=$95V (green triangles) and  $U_{1}=$65V (blue circles), respectively,
were applied.
One can see that using this method, the FWHM of the echo can be changed.
} \label{compression}
\end{figure}

In this section we report on the CRIB experiments. Each of the weak pulses arrives at the sample within the
time of an electrical pulse applied to the electrodes of the
sample (see the inset of figure \ref{compression}), i.e. when the central peak is
broadened.

\begin{figure}[htb]
\includegraphics[width=230pt]{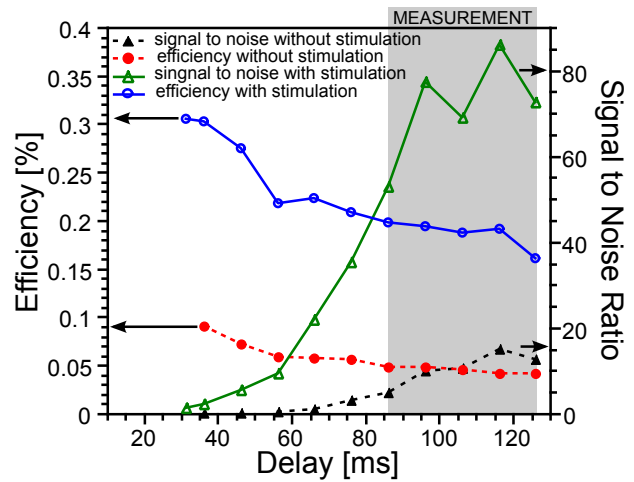}
\caption{(COLOR ONLINE) Efficiency of the memory and signal to
noise ratio as a function of waiting time after state preparation
with and without stimulation laser. For these measurements
$N=2000$ consecutive pulses with a temporal width of 200 nanoseconds were
sent into the sample. The distance between the pulses was 5
microseconds. The number of incident photons per pulse was
$\bar{n}=27$. Note that the signal to noise ratio is a linear function of $\bar{n}$ and the efficiency is independent of the number of input photons in this regime \cite{Lauritzen2010}. The gray box labeled \textit{MEASUREMENT} indicates the time interval in which storage experiments were carried out in the rest of this work.}\label{signal_to_noise}
\end{figure}
We first studied the effect of the fluorescence noise.
In order to do this, we measured the signal to noise ratio as well as the efficiency of the storage, as
a function of the time $T_{wait}$ between the memory preparation
and the beginning of the storage sequence (see fig. \ref{signal_to_noise}). The same
measurements were also performed without the stimulation laser applied.
\begin{figure}[htb]
\includegraphics[width=240pt]{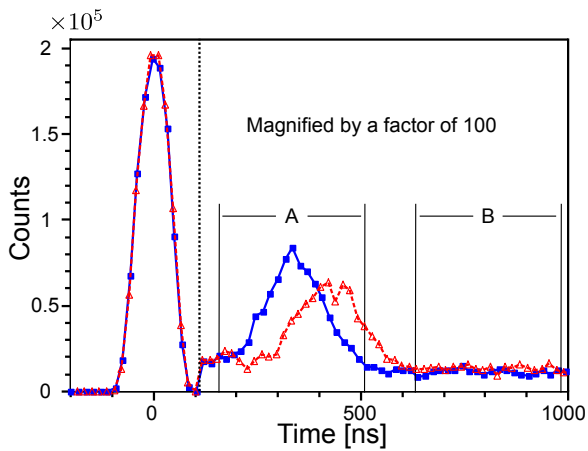}
\caption{(COLOR ONLINE) CRIB echoes for two different switching
times of the electric field. The number of photons per incident
pulse was $\bar{n}=0.9$. Input pulse duration was 100 nanoseconds. The integration time was $2.5\times 10^{4}$ seconds.
Dark counts of 10 Hertz have been subtracted from the data. The voltage
applied on the electrodes was $\pm70$V. The enlargement of the
echo with respect to the incident pulse can be explained by a
spectral mismatch between pulse and absorption line. A and B indicate the time windows used to calculate the signal to noise ratio (see text).}\label{CRIB_echoes}
\end{figure}

To calculate the signal to noise ratio we considered two time windows in the histograms recorded by the TAC (see fig. \ref{CRIB_echoes}).
The first one (A) contains the echo including the noise floor. The second one (B) has the same number of bins and is placed outside the region where the echo is expected, i.e. it only contains the noise floor. From the number of counts $N_{A,B}$ in these windows, the signal to noise ratio can be calculated as
\begin{equation}
 S/N=\frac{N_{A}-N_{B}}{N_{B}}.
\end{equation}

Due to a  better performance of state preparation when the
stimulation laser is applied \cite{Lauritzen2008}, the efficiency
of the memory is significantly higher compared to the case where
it is not applied. At the same time also the noise is reduced, as explained in section \ref{noisereduction}.
However, we noticed a heat input for higher stimulation power. The limited power we could apply to the sample lead to a lower efficiency of the memory. The use of a cryostat with a higher cooling power might thus lead to better results.
Nevertheless, in this work we had to make use of its spontaneous de-excitation in order to lower the
noise rate.

Measurements of storage and retrieval of weak pulses with a mean
number of photons $\bar{n} <1$  using this system have been
reported in \cite{Lauritzen2010}. Fig. \ref{CRIB_echoes} shows
another example with $\bar{n}$ =0.9. Here the input
pulses have a duration of 100 nanoseconds and the initial line is broadened by
applying a voltage of 70 volts to the electrodes which corresponds roughly to a broadening of the central peak by a factor of 3 \cite{Minar2009a}. In that case, the
storage efficiency is of the order of 0.1\%. Higher storage and retrieval
efficiencies have been reported in \cite{Lauritzen2010} for
smaller broadenings and longer pulse durations.

Remarkably, we observe that the efficiency does not decay as a function of $T_{wait}$ to the
extent expected from the ground state Zeeman lifetime.
This can be explained by a non-negligible contribution of long
lived holes \cite{Hastings-Simon2008a} to the spectrally tailored
absorption profile. In order to confirm this we measured the
lifetime of a spectral hole created with stimulation laser
applied. Besides contributions from the excited state decay
($T_{1}=11ms$) and the second ground state Zeeman level decay
($T_{Z}=130ms$) we observed a third decay time of about 15 minutes
(a spectral hole could still be observed 1 hour after optical
pumping).
The origin as well as its dependence on experimental
conditions (such as pump and stimulation power) of the occurrence of these long lived holes so far remain unclear and further
investigations have to be carried out. This would be of great
interest since the results presented in this work only could be
obtained because the persistent hole allowed us to wait ($>85$ms)
for the ions to de-excite without loosing to much in retrieval
efficiency. Another interesting feature of the long-lived holes is
that they in principle allow to work at temperatures above the
limit due to spin-lattice relaxation between the ground-state
Zeeman levels presented in \cite{Hastings-Simon2008a}.
\begin{figure}[htb]
\includegraphics[width=214pt]{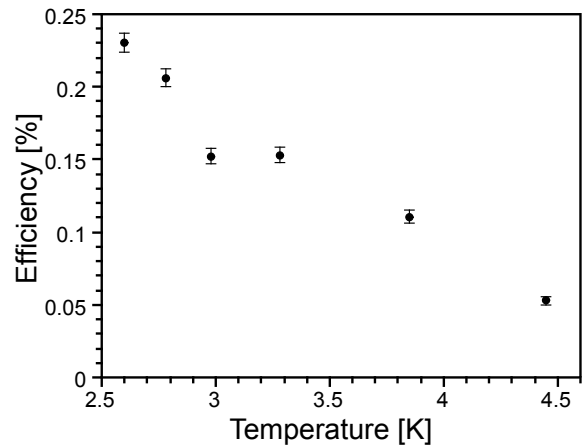}
\caption{Efficiency of the CRIB memory as a function of
temperature. At temperatures significantly above 3 Kelvin there can be
no contribution to the tailored absorption profile from population
transferred between ground state Zeeman levels \cite{Hastings-Simon2008a}. The
existence of very long-lived (persistent) holes and there contribution to the
absorption profile allows the observation of a CRIB echo at
temperatures up to 4.5 Kelvin. The number of incident photons per pulse
was $\bar{n}=10$, integration time for each measurement was 200s.
For the positive and the negative electrical pulses a voltage of
$\pm50$V was applied to the electrodes.} \label{T-dependence}
\end{figure}
Figure \ref{T-dependence} shows the efficiency of the CRIB memory
as a function of temperature.  Above 3.5 kelvin there
is no population transfer between the ground-state Zeeman levels
possible \cite{Hastings-Simon2008a}. We can thus attribute the feasibility of the CRIB
protocol in this regime to the contribution of long
lived holes. The existence of the spectrally tailored absorption
profile at these temperatures was confirmed in a simple absorption
measurement.

These results have allowed us to perform the experiments presented in
\cite{Lauritzen2010}.
At this point we would like to present
further measurements which show how the electric field does not
only allow the storage and retrieval of light, but also provides a
resource for shaping the temporal width of optical pulses.

Let us assume an initial artificially broadened line with a width
$\Gamma_{1}$ and an incident light pulse that spectrally matches
this line. The temporal width of this light pulse is proportional
to the reciprocal of the spectral width
($t_{1}\propto 1/\Gamma_{1}$) since it is simply given by
the Fourier transform. If we now mirror the broadening by
reversing the polarity of the electric field, as described above,
$\delta_{j}\rightarrow -\delta_{j}$ ($\Gamma_{2}=\Gamma_{1}$), the
echo coming out will have the same spectral and thus also temporal
shape as the incident pulse. If we choose the electric fields to
be asymmetric, i.e. $\delta_{j}\rightarrow -\alpha\delta_{j}$ ,
the spectral width of the echo will be changed by a factor
$\alpha$ ($\Gamma_{2}=\alpha\Gamma_{1}$) and so the temporal
width  will change as well. This method was studied in detail by Moiseev and Tittel \cite{Moiseev2010}.
It could serve as a tool for bandwidth matching of
broadband photons to narrowband quantum memories or to increase bitrates in quantum communication networks.
Its feasibility was experimentally demonstrated
in a rubidium vapor using a switchable magnetic field gradient \cite{Hosseini2009}.

Figure \ref{compression} shows the temporal width (full width at half maximum)
for six different combinations of the first and second electrical field ($U_{1}$ and $U_{2}$)
used to induce and reverse the broadening. One can clearly see how
the echo can be compressed or stretched with respect to the input pulse.
Note that for this measurement we were limited by the width of
the initial peak causing a fast decay of the efficiency with
storage time. Temporally long pulses occur truncated
by the envelope given by this decay curve. Nevertheless,
we still can observe the expected effect without correcting for the decay.

\section{Storage of weak coherent pulses using the atomic frequency comb protocol}\label{AFC}
We now describe the measurements of storage and retrieval using
the AFC protocol. Using the spectral hole burning technique
described in Section \ref{spectral}, we created an AFC with
by turning off the pump laser 14 times during one frequency sweep.
We then sent in weak pulses of light of 100 nanoseconds to be mapped on the crystal.

\begin{figure}[htb]
\includegraphics[width=214pt]{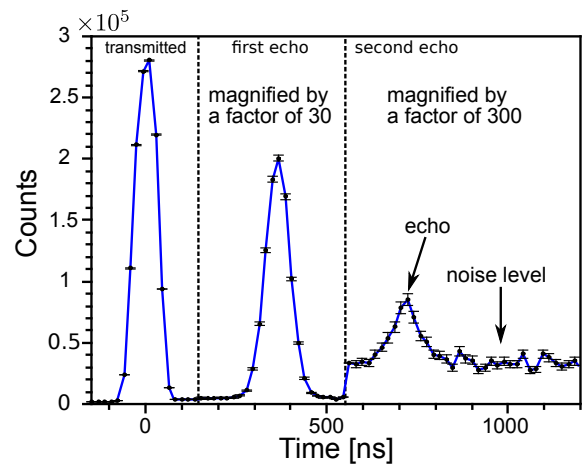}
\caption{(COLOR ONLINE) AFC echoes at the single photon level. The graph shows a
histogram of $8.64\times 10^8$ trials (10h integration time) with a mean
photon number of $\bar{n}=0.5$ per incident pulse. The peak on the
left is the transmitted part of the incident photons. On the right
one can see the echoes at the first and the second time (360ns and
720ns) the atomic dipoles get back into phase.} \label{AFCechoes}
\end{figure}

Figure \ref{AFCechoes} shows the result of a measurement with an input pulse with a mean number of $\bar{n}=0.5$ photons. The comb had a finesse of $F\approx2.6$,
an absorbing background of $d_{0}=1.5\pm0.3$ and an optical depth of the peaks of
the comb of $d=0.5\pm0.2$. One can clearly see the
first echo at a delay of $\frac{1}{\Delta}=360$ns.
The efficiency for this echo is $\eta_{1}=0.7\%$.
As expected from equation \ref{rephasingtime}, a second echo occurs at
$\frac{1}{\Delta}=720$ns since here the atomic dipoles get back
into phase for a second time. Since the storage time is twice as long and
due to the finite width of the peaks \cite{Afzelius2009a}, this echo is
much weaker and almost not visible over the noise level ($\eta_{2}\approx2\times10^{-4}$).
 Note that the finesse of the comb was chosen to reach the maximal efficiency one
can expect for the values $d$ and $d_{0}$ (see Eq. \ref{AFCeq}).
Measurements at different values of finesse confirmed
this expectation. Note that other experiments have already demonstrated much higher efficiencies at other wavelengths in different materials \cite{chaneliere-2009,Amari2010,Sabooni2010,Clausen2010}.

In order to show that the coherence of the light is preserved
during the storage, we performed an interference experiment using
a local oscillator. To observe an interference fringe, we
use the fact that the phase of the AFC echo can be controlled with
the position in frequency of the AFC.
As described in detail in \cite{Afzelius2009a} the occurrence and
decay of the re-emitted light field is given by the Fourier
transform $\tilde{\theta}(t)$ of the periodic atomic distribution
$\theta(\delta)$. At times $t=\frac{m}{\Delta}$ (with
$m\in\mathbb{N}$) after the absorption an echo will be emitted due
to the periodical rephasing of the atomic dipoles. If the central
frequency of the AFC is shifted by $\Delta_{0}$ with respect to
the incident light, i.e. $\theta(\delta)\rightarrow\theta(\delta+\Delta_{0})$, an
additional phase factor occurs in the Fourier transform:
$exp(-i2\pi\Delta_{0}t)\tilde{\theta}(t)$. The echoes will aquire a phase
\begin{equation}
 \phi_{m}=m2\pi\frac{\Delta_{0}}{\Delta}=m\phi_{1}.
\label{phase}\end{equation}
with respect to the case where $\Delta_{0}=0$.
\begin{figure}[htb]
\includegraphics[width=214pt]{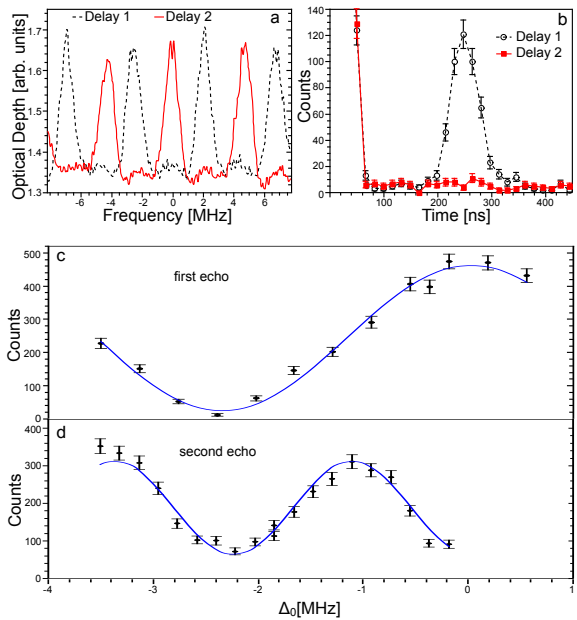}
\caption{(COLOR ONLINE) Results of the interference experiments described in the text. a) Atomic frequency combs (zoom) for two different delays $\tau$ (see figure \ref{pulse_sequence}b) are shifted by $\Delta_{0}$. b) Constructive (delay 1) and destructive (delay 2) interference of the AFC echo with the transmitted part of a second pulse send into the sample for the two cases shown in a. c) Interference fringes:  successively shifting the comb by changing the delay $\tau$ (fig. \ref{pulse_sequence}b) allows to observe a phase shift range of $2\Pi$. The visibility is $89\pm3\%$. d) The same experiment with the second AFC echo. One can see that twice the phase is accumulated. This also explains the lower visibility of $66\pm3\%$ (see text).  Dark counts have been subtracted from the data. $\bar{n} \approx 0.9$ for c and $\bar{n}\approx 9$ for d.}
\label{interference}
\end{figure}

In order to observe the phase $\phi_{1}$
we sent two consecutive pulses into the sample and recorded the
output for different $\Delta_{0}$. Width, amplitude and timing of
the second pulse had been chosen such that its transmitted part,
which serves as local oscillator, perfectly matches the echo of
the first pulse. The position of the comb was changed by changing the trigger delay between the
frequency sweep and the pump sequence (fig. \ref{pulse}
and \ref{interference}a). Figure \ref{interference}b shows the case of constructive
and destructive interference between the echo and the second
pulse. The number of incident
photons for the first pulse was  $\overline{n}\approx0.9$. Note that the average
number of photons per pulse that interfere in this experiment
hence is of the order of $10^{-2}$! Figure
\ref{interference}c shows the number of counts in the echo as a
function of $\Delta_{0}$. The sinusoidal fit gives a visibility of
$89\pm3\%$. Figure \ref{interference}d shows the same
type of interference fringes but for the second AFC-echo. In order
to obtain a signal clearly above the noise the number of incident
photons per pulse was set to $\bar{n}\approx9$. According to equation
\ref{phase}, the acquired phase $\phi_{2}$ is twice as large for a fixed
$\Delta_{0}$. This corresponds well to what we observe. With
$66\pm3\%$ the visibility is significantly lower than for the
first echo. This can be explained by the phase noise induced by
the jitter of the laser used as local oscillator.  Let us assume that during the time between state
preparation and the measurements the laser has shifted by $\delta\Delta_{0}$ due to the jitter.
Introducing it into equation \ref{phase} we get:

\begin{equation}
\phi_{m}=m2\pi\frac{\Delta_{0}+\delta\Delta_{0}}{\Delta}=m(\phi_{1}+\delta\phi_{1}).
\label{phase_equation}
\end{equation}

Assuming a Gaussian distribution  with a width $\sigma$ for the
phase noise, the visibility for $m=1$ is given by
\cite{Minar2008}.
\begin{equation}
 V_{1}=e^{-\sigma^2/2}.
\end{equation}

For $m>1$ the phase noise is m times larger. Therefore
\begin{equation}
 V_{m}=e^{-(m\sigma)^2/2}=V_{1}^{m^{2}}.
\end{equation}
In our case $V_{1}=0.9$. The value of 0.66 we found experimentally
for $V_{2}$ fits well with this explanation.

The results of this experiment confirm that the coherence of the absorbed and re-emitted light is well preserved in an AFC-type light-matter interface. Moreover we see that one can use the AFC itself to shift the phase in a controlled way. This capability provides an interesting tool for interference experiments involving an AFC-storage \cite{Clausen2010}. One can consider the output of this device as one of the outputs of an unbalanced Mach-Zender interferometer. In this picture the transmitted part of the incident light has taken the short, the re-emitted part the long (phase shifted) arm.

\section{Combined AFC-CRIB storage scheme}
\label{combination}

\begin{figure}
\includegraphics[width=214pt]{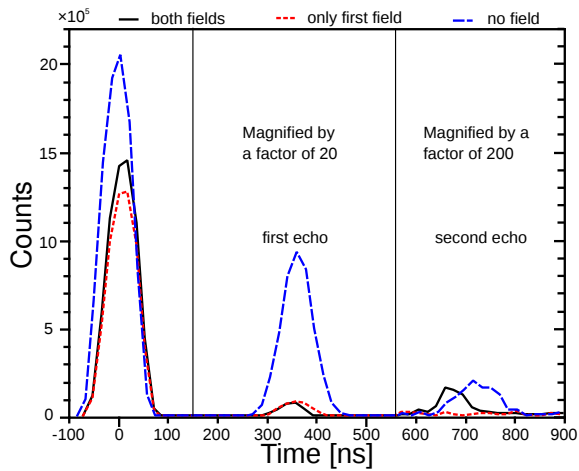}
\caption{(COLOR ONLINE) Here we show control of AFC echoes using the CRIB effect. The dashed line (in blue) is a reference meaurement where no CRIB effect was used. The short-dashed line (red) shows how the first echo can be suppressed by applying an electric field gradient to the sample. In practice a small echo can still be seen since we could not apply a large enough broadening with respect to the comb peak separation. Note that since the applied field is not reversed here, also the second echo is suppressed. The solid line (black) show the results if the broadening is reversed at $t=1/\Delta$, such that the initial coherence is recovered at $t=2/\Delta$, resulting in a recovery of the second echo. Here $\bar{n} \approx 8$ and dark counts have been subtracted from the data.}
\label{AFCCRIB}
\end{figure}

We have here studied two storage schemes having different properties in a two-level configuration. The main advantage of CRIB is the on-demand read-out while for AFC it is the high multimode potential. Yet, the drawback of CRIB is its lower multimode potential whereas for AFC it is the lack of on-demand read-out (predetermined storage time). Note that, in order to turn the AFC into an on-demand memory one can use so-called spin-wave storage \cite{Afzelius2009a,Afzelius2010}, but this would require a three-level configuration with two ground-state spin levels and an optically excited state.

We here propose a modified scheme where we combine the two methods. The applied inhomogeneous broadening in CRIB can be used as a more general tool for controlling the collective emission rate, as we have shown in \cite{Minar2009a}. In particular one can inhibit the revival of the first AFC echo by applying a large enough broadening after the absorption of the input pulse. Let us here remind that the AFC will rephase periodically due to the periodic comb structure. Hence, by reversing the applied inhomogeneous broadening between, for instance, the first and second echo, one can choose to only allow the material to emit the second echo. In Figure \ref{AFCCRIB} we show an example of such a manipulation of AFC echoes using the CRIB technique.

In Figure \ref{AFCCRIB} (and also in Fig. \ref{AFCechoes}) we can also observe that the echo efficiency decreases with the echo number $m$. This can be understood within the AFC theory \cite{Afzelius2009a}, which says that the comb finesse causes an intrinsic dephasing during the time spent in the memory. In the case where all previous $m-1$ echoes before the $m$-th echo have been suppressed using the CRIB effect, the efficiency of the $m$-th echo can be calculated using eq. (\ref{AFCeq}), with the last dephasing factor replaced by $e^{(-m^2/F^2)(\pi^2/4 \ln 2)}$. Otherwise energy will be emitted in the previous echoes and the energy emitted in the $m$-th echo is not trivial to calculate. Let's also note that this result holds if the retrieval efficiency is low, such that a small amount of energy is emitted in each echo (a reasonable assumption in our experiment). The above arguments lead to the conclusion that in order for higher order echoes to be efficient, the comb finesse should be increased. Then this technique can be used to bias the retrieval efficiency towards higher order echoes.

To summarize, by combining the two techniques one can create an AFC memory where the emission time can be controlled in discrete steps, a digital quantum memory. This while retaining it's high multimode potential. Note that experimental results similar to those presented here have been obtained at the University of Calgary \cite{Tittel2010}.

\section{Possible improvements}

The main issue that occurs when working with an erbium doped crystal as material for a quantum memory lies in the inefficient optical pumping. As described in sections \ref{experimental} and \ref{CRIB}, this causes an absorbing background, limits the achievable optical depth of the prepared absorption feature and leads to fluorescence noise. As shown in \cite{Lauritzen2008}, the pumping efficiency can be further increased by mixing the spins of the excited state using radio-frequency (RF) waves. In this way one improves the effective branching ratio and thus the probability for a population transfer.
However, the application of the RF wave causes a heat input into the sample which can lower the lifetime of the ground state \cite{Hastings-Simon2008}. This technical problem can be easily solved with a sufficient cooling power of the cryostat used for the experiments. However, for the measurements presented in this work, such a cryostat was not at our disposal.
The limited cooling power also put a bound to the stimulation power we could apply to our sample. A higher stimulation power would reduce the fluorescence noise even more and increase the efficiency of the population transfer at the same time. However, we noticed a measurable heat input by the stimulation laser into the sample at high power. Also here a higher cooling power as well as a better thermalization of the sample  would probably allow a higher efficiency and a lower noise level and thus to a much better signal to noise ratio.
It is likely that the fluorescence noise has its origin also in decay channels involving other ground state (crystal field) levels than the
ones used for the experiments. The signal to noise ratio could thus be further improved by spectral and polarization filtering after the sample.

The most promising feature that played a role are the persistent holes mentioned in \ref{CRIB}. As already discussed, their origin so far remains unclear. We thus suggest to perform  a detailed analysis of their origin. If they are due to another level not being either of the two ground state Zeeman levels used in the experiment, which can be populated in a controlled way, one might use this level as shelving state. A third ground state level with a long life- and coherence time would allow one to extend both protocols with a spin wave storage. This would be the enabling feature for AFC- or CRIB-memory with spin wave storage at telecommunication wavelengths.
Additionally we suggest to also explore different host materials \cite{Sun2002} and to study the influence of the direction and strength of the external magnetic field in more detail.
Storage time, both for the AFC and the CRIB, can be increased by creating a narrower initial spectral feature. This would in turn require a frequency stabilized laser.

\section{Conclusions}
We have presented experimental results obtained with two different approaches for a light-matter interface at the single photon level.
By using an erbium doped doped into an yttrium ortho silicate crystal (Er$^{3+}$:Y$_{2}$SiO$_{5}$), we were able to carry out these measurements at telecommunication wavelengths.
The main difficulty when working with Er$^{3+}$:Y$_{2}$SiO$_{5}$ is the inefficient optical pumping.
Fluorescence noise from ions left in the excited state after state preparation is an issue when working at the single photon level in this material. However, we have shown how to minimize this source of noise.
We have presented results concerning the possibility of pulse compression and stretching with the CRIB scheme. We have also shown that the presence of persistent holes not only allows one to perform measurements at the single photon level at a reasonable noise level, but also to observe CRIB echoes significantly above the temperature expected from measurements of the Zeeman level lifetime in this material \cite{Hastings-Simon2008a}. However, the efficiency of the optical pumping in this material remains low and further research on this is required in order to make the device a useful tool for future experiments in quantum information science.

Furthermore we have presented AFC experiments at the single photon level in the same material.
We have demonstrated the preservation of coherence during storage by an interference experiment using a local oscillator. These measurements also show how the phase of the echo can be changed with respect to the input pulse by a frequency shift of the AFC.
The preservation of coherence for the second AFC echo was also  shown. We observe that its phase is shifted twice as much as the phase of the first echo. The limited visibility of the interference fringes of both measurements could be explained by the phase noise of the local oscillator.


Overall we believe that our experiments show the potential of erbium doped solids for quantum memory applications and we hope that it will motivate some further material research.

\begin{acknowledgments} We would like to thank Claudio Barreiro for technical assistance.  Furthermore would we like to thank Christoph Simon, Nicolas Bruno Sangouard and Imam Usmani for useful discussions. This work was supported by the Swiss NCCR Quantum Photonics, by the European Commission under the Integrated Project Qubit Applications (QAP) and the ERC-AG Qore.
\end{acknowledgments}

\end{document}